\documentstyle[aps,pra,multicol,epsfig,array]{revtex}

\newcommand \be{\begin{eqnarray}}
\newcommand \ee{\end{eqnarray}}

\draft

\begin{document}

\title{Branching of negative streamers in free flight}

\author{Andrea Rocco$^1$, Ute Ebert$^{1,2}$, and Willem Hundsdorfer$^1$}
\address{$^1$ Centrum voor Wiskunde en Informatica,
P.O.Box 94079, 1090 GB Amsterdam, The Netherlands\\
$^2$ Dept. Physics, TU Eindhoven, 5600 MB Eindhoven, The Netherlands}
\date{\today}

\maketitle

\begin{abstract}
%We analyze the minimal continuum model for negative streamers 
%in a non-ionized and non-attaching gas 
%with local impact ionization reaction.
We recently have shown 
%by a combination of simulations and analytical insight 
that a negative streamer in a sufficiently high homogeneous field 
can branch spontaneously due to a Laplacian instability, rather than
approach a stationary mode of propagation with fixed radius. 
In our previous simulations, the streamer started from a wide
initial ionization seed on the cathode. We here demonstrate
in improved simulations that a streamer emerging from a
single electron branches in the same way.
In fact, though the evolving streamer is much more narrow,
it branches after an even shorter propagation distance.
\end{abstract}

\begin{multicols}{2}

Streamers are widely known phenomena in electric breakdown:
long and rapidly growing channels of high ionization penetrating
a region of low or vanishing ionization under the influence
of a strong electric field \cite{Raether,Rai}.
In experiments, streamers frequently are seen to branch.
On a phenomenological level, the so-called dielectric break-down
model (DBM) \cite{DBM,DBM2,DBM3}, a variation of diffusion limited aggregation
\cite{DLA}, has been suggested as a stochastic process explaining 
the branching. Such models are inspired by earlier concepts of 
streamer propagation where the space charge was assumed to be smeared 
out over the full streamer head. Branching then would occur due to
randomly distributed ionization avalanches around the streamer.
Such stochastic phenomenological models on larger scale have not 
been related to particular discharge models.

However, simulations \cite{DW,Vit} have established 
that the space charge of the streamer is concentrated in a thin layer 
around the head rather than being smeared out over the full head. This leads
to a different field distribution and a much faster propagation mode, 
and to a revival of the concept of an ideally conducting streamer 
formulated by Lozansky and Firsov \cite{Firsov}. In particular,
this charge distribution allows for a completely different ``Laplacian''
instability mechanism of streamers \cite{Ute,Sit} that can operate 
even without any randomness quite like in viscous 
fingering, dendritic growth of solids into undercooled melts etc. 
It naturally emerges in the minimal streamer model (\ref{1}) -- (\ref{5}). 
In a recent paper \cite{PRLMan}, we found that in contrast 
to previous expectations, 
this Laplacian instability mechanism can cause the spontaneous 
branching of streamers propagating in a strong homogeneous field. 
The system was identical to that of previous simulations \cite{DW,Vit}
except that the field was twice as high.
%
%We as well as others \cite{Nat} interpreted our results \cite{PRLMan} 
%as the prediction from an explicit discharge model that a streamer 
%can branch in a homogeneous field. 
However, our simulation results were 
somewhat limited by numerical constraints. Because of the unexpected 
results \cite{Nat} and some non-smooth structures in the figures, 
some researchers wondered whether the figures showed a numerical
rather than a physical instability \cite{Focus}.

Therefore we here present new simulations with improved numerics,
and we show more details of the evolution.
In particular, we increased the number of grid points in the simulations
from 1000$\times$1000 to 2000$\times$2000, which improves both the
accuracy of the numerical output data and the quality of the contour plots,
so the new plots are smooth. Furthermore, we changed the boundary 
condition on the cathode from homogeneous Neumann $\partial_z\sigma=0$
to homogeneous Dirichlet $\sigma=0$
for the electron density $\sigma$. This means that while previously
the electrons freely could flow from the metal of the electrode
into the gas, this current is now suppressed. With these improvements,
the choice of the initial conditions
now ceases to be constrained by numerical considerations.
We continue to use cylinder symmetry to calculate effectively on 
a 2-dimensional grid. Doing so, we keep relying on the analytical 
argument that the constraint of cylindrical geometry suppresses 
some instability modes, and that therefore a truly 3-dimensional 
system would become unstable earlier or at the same time as the 
system with symmetry constraint, but certainly not later.

The improvements of the numerical code allow now for the simulation 
of a streamer starting from a single electron on the cathode rather 
than from a wide initial seed as in \cite{PRLMan}.
Our new figures show the streamer branch in free flight. 
The initial electron on the cathode creates first an avalanche, 
then a streamer, and finally the streamer splits. The situation 
resembles the historical experiments of Raether \cite{Raether}, 
except that the field is higher.

In detail, we investigate the minimal streamer model,
i.e., a ``fluid approximation'' with local field-dependent
impact ionization reaction in Townsend approximation \cite{Rai}
in a non-attaching and non-ionized gas. In dimensionless units 
\cite{Ute,Sit,PRLMan}, the model has the form: 
\begin{eqnarray}
\label{1}
\partial_t\;\sigma \;-\; 
\nabla\cdot\left(\sigma\;{\bf E} + D\;\nabla\sigma\right)
&=& \sigma \; f(|{\bf E}|)~,
\\
\label{2}
\partial_t\;\rho \;
&=& \sigma \; f(|{\bf E}|)~,
\\
\label{3}
\rho - \sigma &=& \nabla\cdot{\bf E}~~,~~{\bf E}=-\nabla \Phi~,
\\
\label{5}
f(|{\bf E}|)&=&|{\bf E}|\;e^{-1/|{\bf E}|}~.
\end{eqnarray}
Here $\sigma$ and $\rho$ are the densities of electrons and
positive ions, ${\bf E}$ is the electric field, $\Phi$ the electric
potential and $D$ the dimensionless diffusion constant.
The mobility of ions is neglected. The same model was investigated 
in \cite{DW,Vit}.
The translation to physical units depends on the type and the pressure $p$
of the gas. With an effective field-dependent impact ionization
coefficient $\alpha(E)=Ap\;\exp\left(-Bp/|E|\right)$ and electron mobility
$\mu_e=\bar\mu/p$ as in \cite{Rai}, and with the parameter values
for nitrogen as in \cite{DW,Vit}, we get for the scales
of length, time, field and particle density 
\be
\label{scale}
&&l_0 \simeq \frac{2.3 \;\mu \mbox{m}}{(p/1{\rm bar})}~~,~~
t_0 \simeq \frac{3 \;\mbox{ps}}{(p/1{\rm bar})}~~,~~
E_0 \simeq 200 \;\frac{\mbox{kV}}{\mbox{cm}} \;\frac{p}{1{\rm bar}}~,
\nonumber\\
&& n_0\simeq \frac{5\cdot 10^{14}}{\mbox{cm}^3}\;
\left(\frac{p}{1{\rm bar}}\right)^2 
\simeq 2\cdot 10^{-5}\cdot [N_2]\cdot \frac{p}{1{\rm bar}}~,
\ee
where $[N_2]$ is the neutral gas particle density.
The pressure dependent scaling relations are of interest
for laboratory experiments, as well as for so-called sprite discharges 
\cite{Pasko} in the mesosphere above thunderclouds.

The simulation is performed with the same voltage of 1000 and electrode
distance of 2000 as in \cite{PRLMan}, corresponding to a field
of 100 kV/cm and an electrode distance of 4.6 mm for nitrogen 
at $p=1$ bar. As an initial condition, one electron 
was distributed over one cell of the numerical grid next to 
the cathode. Because of the diffusive broadening of the 
drifting avalanche, this ``distribution'' of the initial
electron is permissible.
%\be 
%\sigma({\bf r},t=0)=10^{-4}\;e^{-{\bf r}^2}~.
%\ee

In the figures, we omit the avalanche phase and concentrate on
the developed streamer and its branching at times $t=500$,
525, 550 and 575. The field is applied in the $z$ direction,
and $r$ is the radial coordinate extending up to $r=2000$ to suppress
artifacts from the lateral boundaries.
The simulation shows a streamer not connected to 
the electrode. It has the conical shape of the initial avalanche
created by a single electron in a homogeneous field as first
found in Raether's experiments \cite{Raether}.

\begin{figure}
\centerline{{\psfig{figure=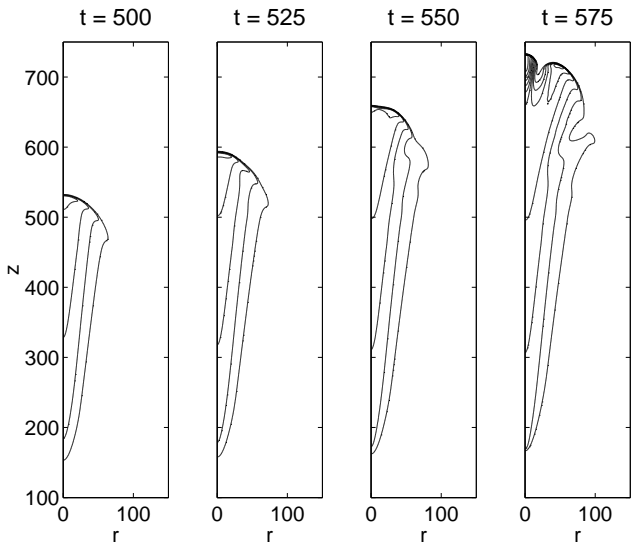,height=7.3cm}}}
\caption{Evolution of an anode directed 
streamer in a strong homogeneous background field. 
The planar cathode is located at $z=0$ and the planar anode at $z=2000$. 
The radial coordinate extends from the origin up to $r=2000$ 
to assure homogeneous field conditions. Shown is $100\le z\le750$ 
and $0\le r\le150$ with equal axis scaling. The thin lines denote 
levels of equal electron density $\sigma$ with increments of 0.15.}
\end{figure}

Fig.~1 shows the electron density $\sigma$ within the whole streamer body
at four snapshots of the evolution. Due to the no-current-boundary
condition on the electrodes, the streamer is not connected to the cathode
at $z=0$. The equidensity lines mark multiples of 0.15 in all snapshots. 
The electrodes at $z=0$ and $z=2000$ are outside the region shown.
The qualitative evolution is the same as in our previous paper \cite{PRLMan}:
while the streamer extends, it becomes wider, the electron density increases,
and finally the instability develops.

\begin{figure}[h]  
\centerline{{\psfig{figure=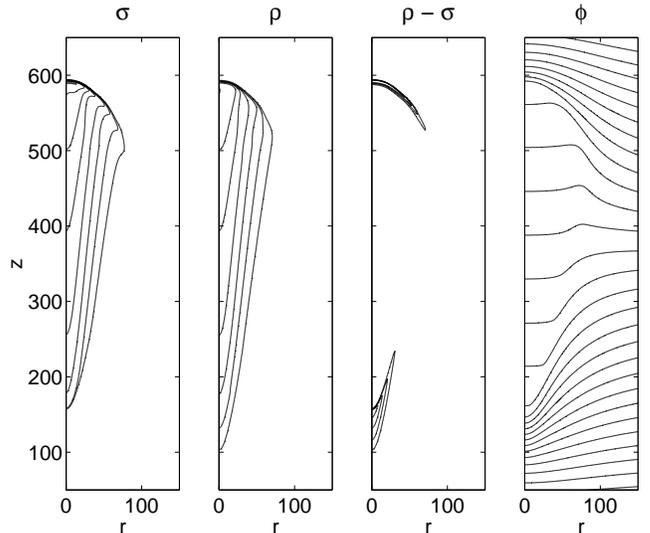,height=7.3cm}}}
\caption{The streamer at time $t=525$. The figures show contour
plots with equal aspect ratio for the electron density $\sigma$, 
ion density $\rho$,
space charge density $\rho-\sigma$ and electric potential $\Phi$.
The level lines for the densities $\sigma$ and $\rho$ are multiples of 0.1,
for the space charge density $\rho-\sigma$ of 0.05, 
and the increment between different levels of the potential
is 10.}
\end{figure}

Fig.~2 shows four different observables of the whole streamer
at one time instant $t=525$. We plot
the electron density $\sigma$, the ion density $\rho$, the space
charge density $\rho - \sigma$, and the electric potential $\Phi$. 
In the propagating front at the tip of the streamer, an overshoot 
of drifting electrons creates a negatively charged layer, while 
the back of the streamer is depleted from electrons and positively charged 
by the essentially  immobile ions left behind. Therefore the 
``ion streamer'' is longer in the back, and the ion density has no overshoot 
in the propagating tip. The resulting space charge densities are shown 
in the third plot: a positive space charge with a maximum density of
$0.25$ in the back, and a negative space charge 
with a minimum density of $-0.25$ in the propagating tip.
While the whole object is electrically neutral, the electric 
polarization leads to a suppression of the field in the interior.
This can be read from the large distance between equipotential lines
inside the ionized body visible in the last plot.
The streamer is indeed approaching the Lozansky-Firsov limit 
of ``ideal conductivity'' \cite{Firsov}.      

Fig.~3 shows a zoom into the tip of the streamer at the same time
steps as in Fig.~1. Again the electron densities $\sigma$ are plotted,
and addionally the equipotential lines $\Phi$. Already at time
$t=525$, the onset of the instability can be concluded from the
somewhat irregular shape of the level lines. The curvature
of the tip decreases, the densities increase, the field in the interior
decreases and the field ahead of the streamer increases.
Consequently, the tip becomes unstable and splits.

\end{multicols}

\begin{figure}[h]  
\centerline{{\psfig{figure=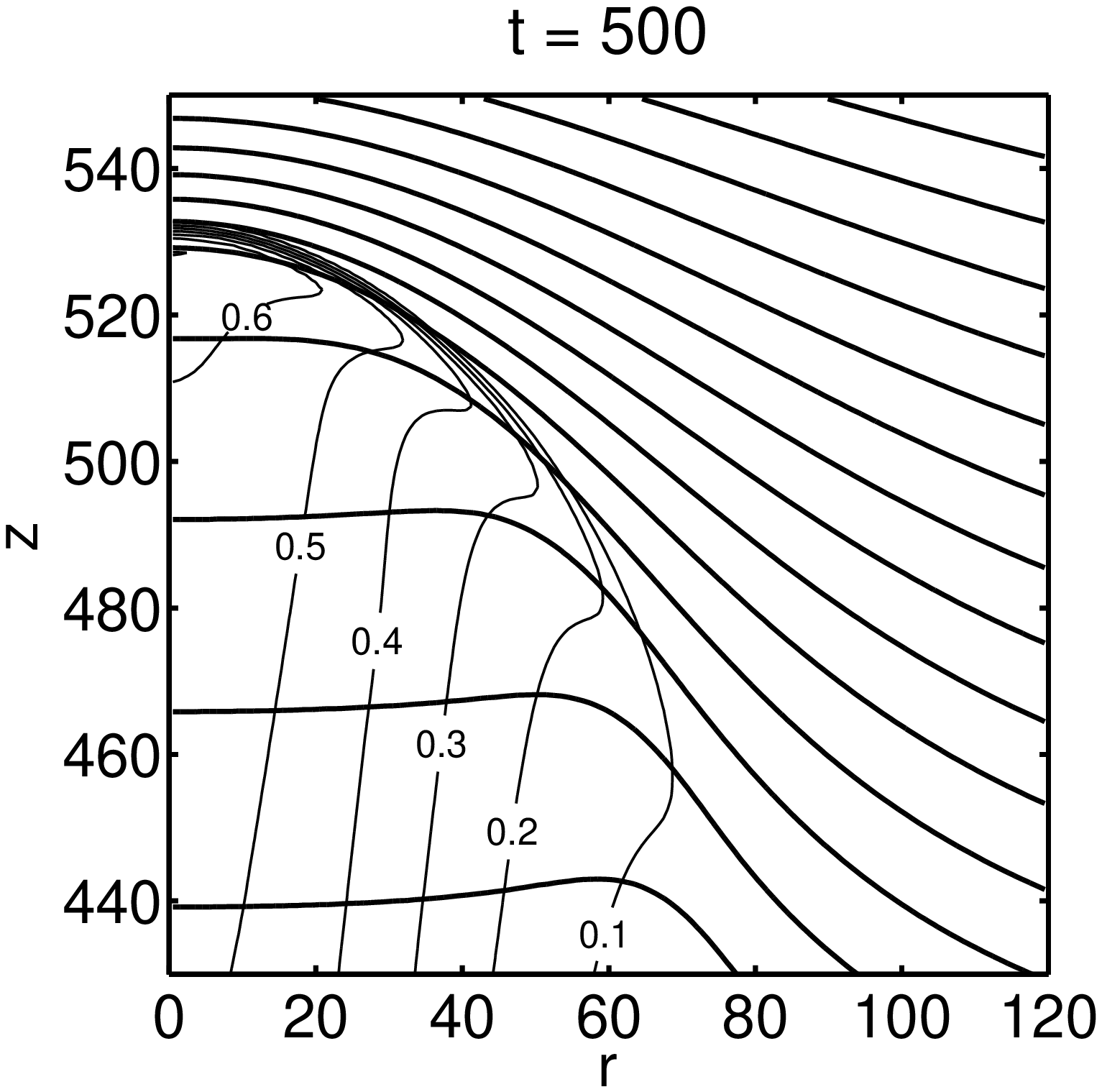,height=7.5cm}}
{\psfig{figure=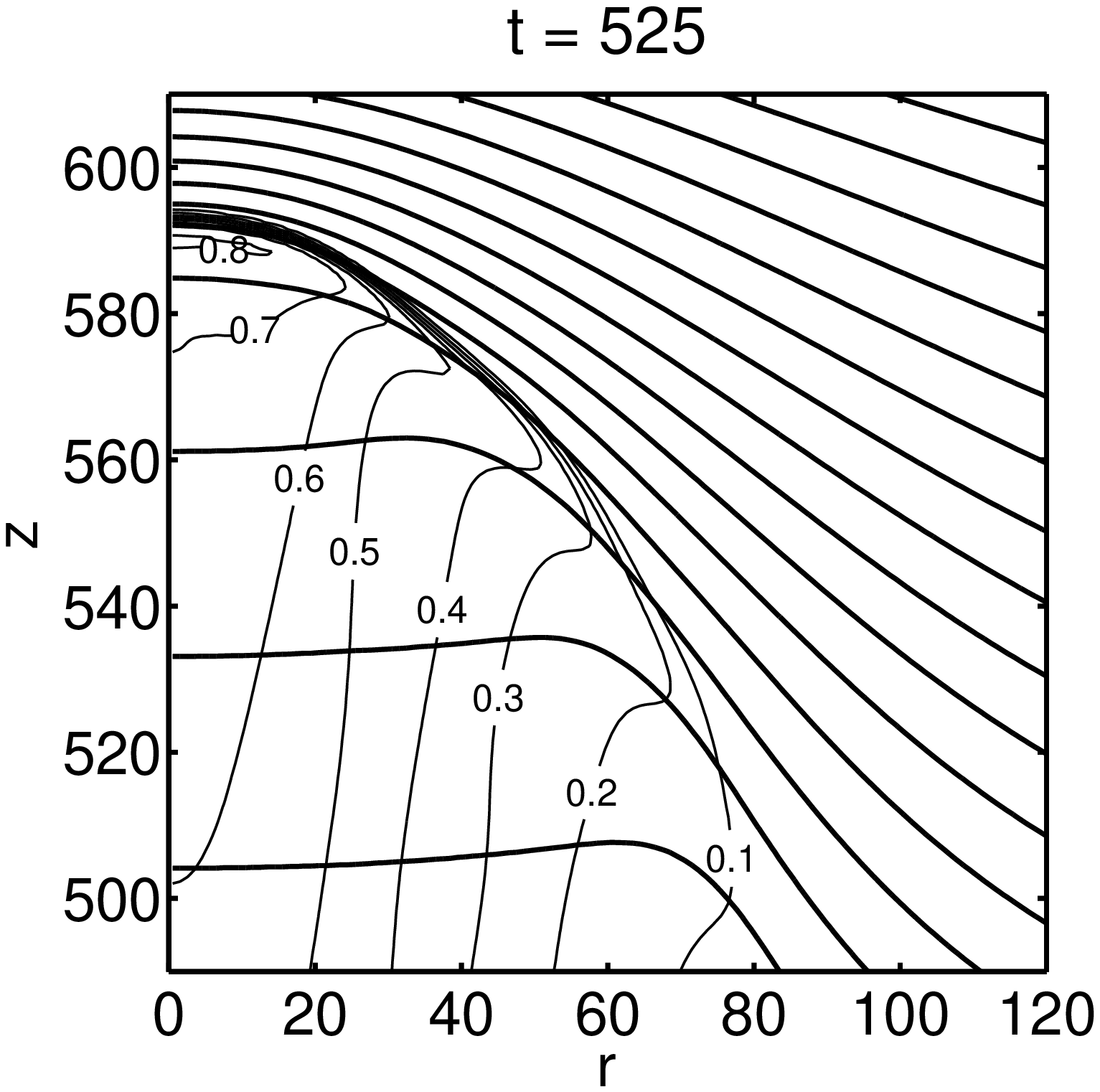,height=7.5cm}}}
\centerline{{\psfig{figure=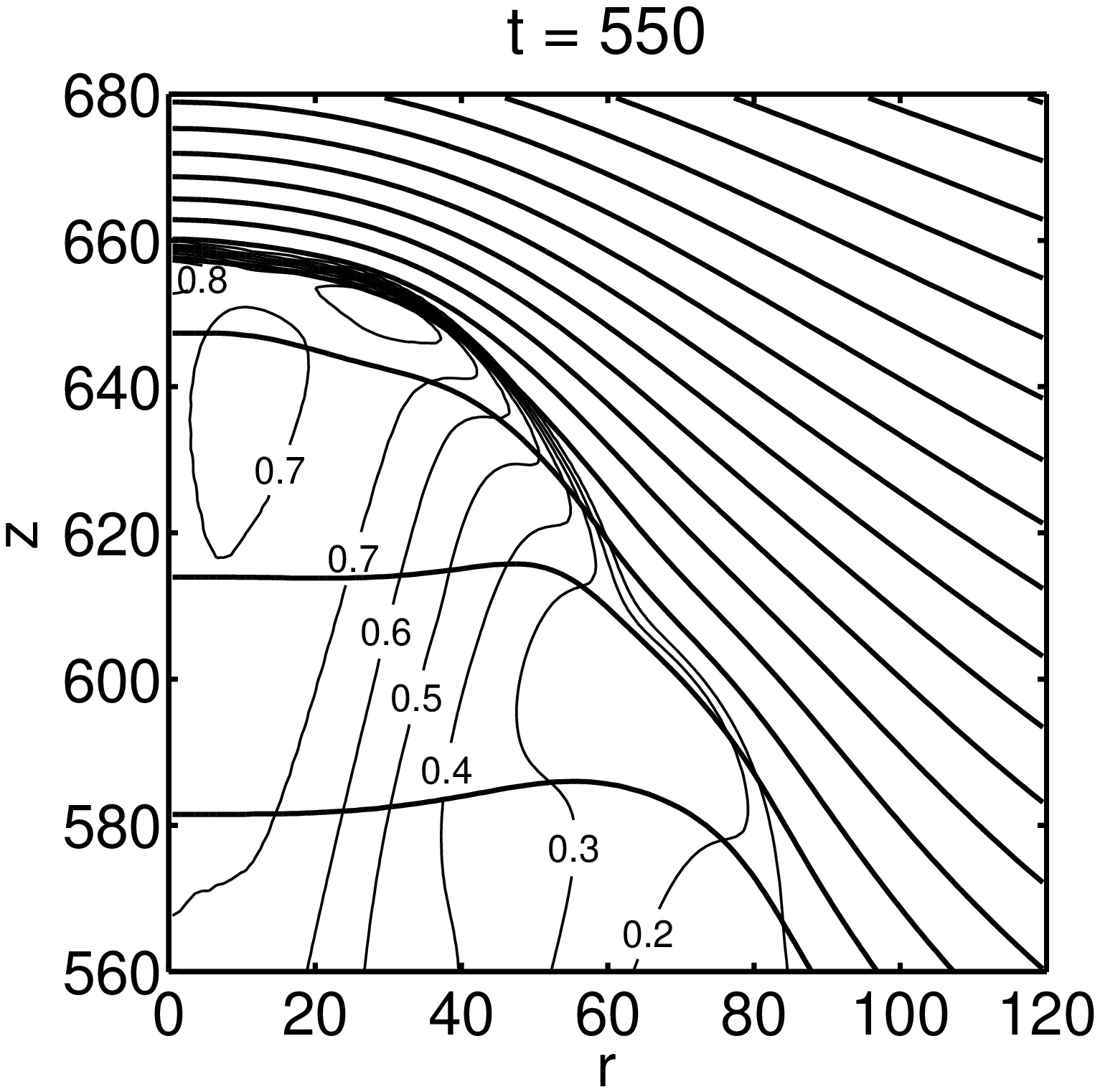,height=7.5cm}}
{\psfig{figure=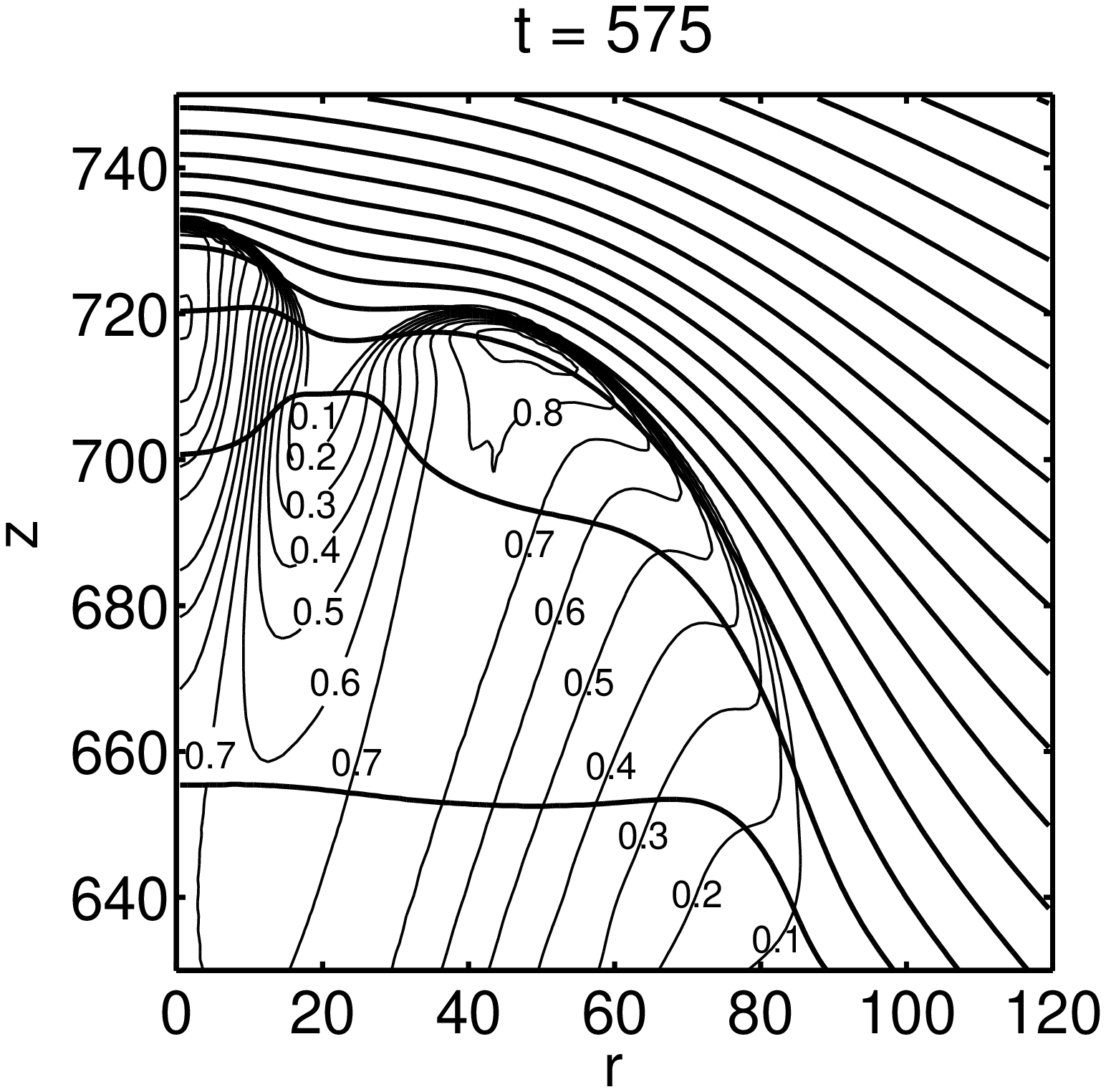,height=7.5cm}}}
\caption{Zoom into the evolution of the streamer tip showing electron 
density $\sigma$ with increments of 0.1 as thin lines and electric 
potential $\Phi$ with increments of 5 as thick lines.
At the last time step $t=575$, the maximal electron density 
in the core of the finger on the axis is 1.5.
All figures have equal aspect ratio and equal axis scaling.
%Between time steps 500 and 525, the ionization front becomes steeper,
%less curved and more equipotential. This creates a state with 
%Laplacian instability. Consequently, tip splitting evolves in
%the later stages.
}
\end{figure}

\begin{multicols}{2}

We conclude with a few remarks: \\
1. It is quite remarkable that the finger of the new
simulations branches after an even shorter travel distance than 
the one growing out of a wide, highly ionized seed in \cite{PRLMan},
though it is more narrow during its whole evolution.\\
2. When the 
initial electron is not placed on the electrode but somewhere in the gap,
simulations show the same figures as in this paper, essentially unchanged 
except for a trivial shift in space.\\
3. The transition times from avalanche to streamer and then further
from streamer into the instability depend quite nonlinearly on the
externally applied field as we will discuss in a forthcoming paper; 
therefore the question whether branching also would occur in a lower 
field after a longer propagation distance, is open.\\
4. Finally, the discovery of sprite discharges \cite{Pasko} 
in the mesosphere gives a new impetus to the study of negative
streamers in high fields, since a negative sprite discharge
propagates into increasing heights, i.e., areas of decreasing
pressure; the scaling laws (\ref{scale}) therefore imply that
the effective field continues to grow along their path.
\\

{\bf Acknowledgement:} A.R.\ was supported by the Dutch Research School 
CPS, the Dutch Physics Funding Agency FOM, and CWI Amsterdam.

\end{multicols}

%\vspace{-0.8cm}

%\begin{figure}[h]  
%\centerline{{\psfig{figure=500new.eps,height=6cm}}}
%\end{figure}

%\vspace*{-0.1cm}

\newpage

%\begin{figure}
%\label{fig1}
%\begin{center}
%\epsfig{figure=final4.eps,width=1.\linewidth}       
%\caption[]{}
%\end{center}
%\end{figure}

\end{document}